\documentclass[oribibl]{llncs}

\usepackage{amssymb}
\usepackage{graphicx}
\usepackage{comment}
\usepackage{amsmath}
\usepackage{color}
\usepackage{url}
\usepackage{pdfsync}
\usepackage{algpseudocode}
\usepackage{algorithm}
\usepackage{wrapfig}

\newcommand{\mc}[1]{\mathcal{ #1}}
\newcommand{\nit}[1]{{\it #1}}
\newcommand{\boxtheorem}{\hfill $\Box$\vspace*{0.2cm}}
\newcommand{\ignore}[1]{}

{\vskip \abovedisplayskip \refstepcounter{lemmaA-counter}%
\noindent {\bf Lemma A.\arabic{lemmaA-counter}.}}%

\newcounter{lemmaA-counter}

    \newcommand{\da}{{\em Datalog}~}
    \newcommand{\dpm}{{\em Datalog}$^\pm$~}
    \newcommand{\de}{{\em Datalog}$^\exists$~}
\begin{document}
\thispagestyle{empty}
\pagestyle{plain}

\mainmatter  

\title{\vspace*{-10mm}Tractable Query Answering and Optimization for Extensions of Weakly-Sticky Datalog$\pm$
\vspace{-4mm}}

\author{{\bf Mostafa Milani} \and {\bf Leopoldo Bertossi}}

\institute{\vspace{-3mm}Carleton University, \ School of Computer Science, \ Ottawa, \ Canada\\
\email{\{mmilani,bertossi\}@scs.carleton.ca}\vspace{-4mm}}


%


%
%

\maketitle


\paragraph{\bf Summary.} \ We consider a semantic class, {\em weakly-chase-sticky} (WChS), and  a syntactic  subclass, {\em jointly-weakly-sticky} (JWS),  of \dpm programs. Both extend that of weakly-sticky (WS) programs, which appear in
our applications to data quality. For WChS programs we propose a practical, polynomial-time query answering  algorithm (QAA). We establish that the two classes are closed under magic-sets rewritings. As a consequence, QAA can be applied to the  optimized programs. QAA takes as inputs the program (including the query) and semantic information about the ``finiteness" of predicate positions.  For the syntactic subclasses JWS and WS of WChS, this additional information is computable.

\vspace{-2mm}
\paragraph{\bf \dpm\!\!.} \ {\em Datalog}, a rule-based language  for query and view-definition in relational databases  \cite{ceri}, is not expressive enough to logically represent interesting and useful ontologies, at least of the kind needed to specify conceptual data models.  \dpm extends \da by allowing
existentially quantified variables in rule heads ($\exists$-variables), equality atoms in rule heads, and program constraints \cite{AC09}.
Hence the ``$+$" in \dpm, while the ``$-$" reflects syntactic restrictions on programs, for better computational properties.

A typical \dpm program, $\mc{P}$, is a finite set of rules, $\Sigma \cup E \cup N$, and an extensional database (finite set of {\em facts}), $D$.
The rules in $\Sigma$ are  {\em tuple-generating-dependencies} ({\em tgds}) of the form $\exists \bar{x}\!P(\bar{x},\bar{x}') \leftarrow P_1(\bar{x}_1), \ldots, P_n(\bar{x}_n)$, where $\bar{x}' \subseteq
\bigcup \bar{x}_i$, and $\bar{x}$ can be empty. $E$ is a set of {\em equality-generating-dependencies} ({\em egds}) of the form $x = x' \leftarrow P_1(\bar{x}_1), \ldots, P_n(\bar{x}_n)$, with $\{x,x'\} \subseteq \bigcup \bar{x}_i$. Finally, $N$ contains {\em negative constraints} of the form $\bot \leftarrow P_1(\bar{x}_1), \ldots, P_n(\bar{x}_n)$, where $\bot$ is false.
\begin{example} \label{ex:program} The following \dpm program shows a tgd, an egd, and a negative constraint, in this order:
$\exists x\;{\it Assist}(y,x) \leftarrow {\it Doctor}(y)$; \ \ $x=x' \leftarrow {\it Assist}(y,x), \ {\it Assist}(y,x')$; \ \
$\bot \leftarrow {\it Specialist}(y,x,z), \ {\it Nurse}(y,z)$.
\boxtheorem
\end{example}
\vspace{-3mm}Below, when we refer to a class of \dpm programs, we  consider only $\Sigma$, the tgds.
Due to different syntactic restrictions, \dpm can be seen as a class of sublanguages
of \de\!\!, which is the extension of \da with tgds with $\exists$-variables \cite{LE11}.

The rules of a \dpm program can be seen as an ontology $\mc{O}$ on top of  $D$, which can be {\em incomplete}. $\mc{O}$ plays the role of: (a) a ``query layer" for $D$, providing  ontology-based data access (OBDA) \cite{lenzerini12}, and (b) the specification of
a completion of $D$, usually carried out through the {\em chase} mechanism that, starting  from $D$, iteratively enforces the rules in $\Sigma$, generating new tuples. This  leads to a possibly infinite
instance extending $D$, denoted with $\nit{chase}(\Sigma, D)$.

The  answers to a conjunctive query $\mc{Q}(\bar{x})$ from $D$ wrt.  $\Sigma$ is a sequence of constants $\bar{a}$, such that
$\Sigma \cup D \models \mc{Q}(\bar{a})$ (or $\nit{yes}$ or $\nit{no}$ in case $\mc{Q}$ is boolean). The answers can be obtained by querying as usual the {\em universal} instance $\nit{chase}(\Sigma, D)$.
The chase may be infinite, which leads, in some cases, to undecidability of query answering \cite{JO84}. However, in some cases where the chase is infinite, query answering (QA) is still
computable (decidable), and even tractable in the size of $D$. Syntactic classes  of \dpm programs with tractable QA have been identified and investigated, among them: {\em sticky}~\cite{AC12,tods14}, and {\em weakly-sticky} \cite{AC12} \dpm programs.

\vspace{-2.5mm}
\paragraph{\bf Our Need for QA Optimization.} \ In our work, we concentrate on the {\em stickiness} and {\em weak-stickiness} properties, because these programs appear in our applications to quality data specification and extraction
\cite{desweb}, with the latter task accomplished through QA, which becomes  crucial.

Sticky programs~\cite{AC12} satisfy a syntactic restriction on the multiple occurrences of variables (joins) in the body of a {\it tgd}.
Weakly-sticky (WS) programs form a class that extends that of sticky programs \cite{AC12}. WS-\dpm  is more expressive than sticky \dpm\!\!, and results from applying the notion of
{\em weak-acyclicity} (WA) as found in data exchange  \cite{FG03}, to relax acyclicity conditions on stickiness. More precisely, in comparison with sticky programs, WS programs require a milder condition on join variables, which is based on a program's {\em dependency graph} and the positions in it with finite rank~\cite{FG03}.\footnote{A position refers to a predicate attribute, e.g. ${\it Nurse}[2]$.}

For QA, sticky programs enjoy {\em first-order rewritability} \cite{tods14}, i.e. a conjunctive query $\mc{Q}$ posed to $\Sigma \cup D$ can be rewritten into a new first-order (FO) query
$\mc{Q}'$, and correctly answered by posing  $\mc{Q}'$ to $D$, and answering
as usual.  For WS programs, QA is $\nit{PTIME}$-complete in data, but the polynomial-time algorithm provided for the proof in ~\cite{AC12} is not  a practical one.

\vspace{-2.5mm}
\paragraph{\bf Stickiness of the Chase.}
 \ In addition to (syntactic) stickiness, there is  a ``semantic" property of programs, which is relative to the chase (and the data, $D$), and is called ``chase-stickiness" (ChS).
 Stickiness implies semantic stickiness  (but not necessarily the other way around)  \cite{AC12}. \ For chase-sticky programs, QA is tractable \cite{AC12}.

Intuitively, a program has the chase-stickiness  property if, due to the application of  a tgd $\sigma$: When a value replaces a repeated variable in the body of a rule, then that value also appears in all the
head atoms obtained through the iterative enforcement of applicable rules that starts with $\sigma$. So, that value is propagated all the way down through all the possible subsequent steps.

\vspace{-7mm}
\begin{figure}[h]
\begin{center}
\includegraphics[width=3.25cm]{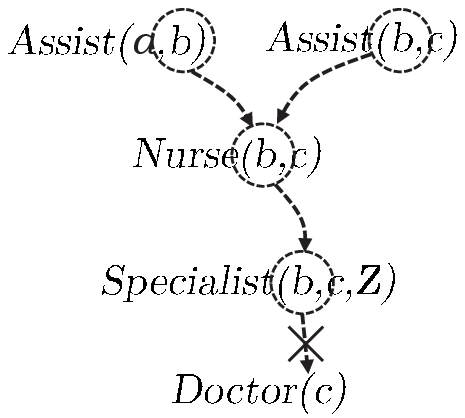}
\includegraphics[width=3.25cm]{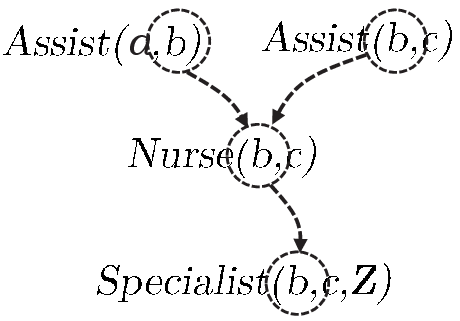}
\end{center}
\vspace{-0.7cm}
\caption{The chase for a non-ChS program and the chase for a ChS program, resp.}
\label{fig:chase}
\vspace{-0.8cm}
\end{figure}

\begin{example}\label{exp:chs} Consider $D=\{{\nit Assist}(a,b),{\nit Assist}(b,c)\}$, and the following set, $\Sigma_1$, of tgds: \
${\it Nurse}(y,z)\leftarrow {\it Assist}(x,y),{\it Assist}(y,z)$; \ \
$\exists z\;{\it Specialist}(x,y,z)\leftarrow {\it Nurse}(x,y)$; \ \
${\nit Doctor}(y)\leftarrow {\nit Specialist}(x,y,z)$. \
 $\Sigma_1$ is not ChS, as the chase  on the LHS of Figure\ref{fig:chase} shows: value $b$ is not propagated all the way down to $\nit{Doctor}(c)$.
However,  program $\Sigma_2$, which is $\Sigma_1$ without its third rule, is ChS, as shown on the RHS of Figure\ref{fig:chase}.
\boxtheorem\end{example}

\begin{figure}[t]
\begin{center}
\includegraphics[width=7.5cm]{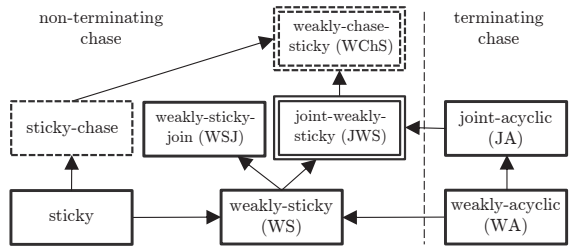}
\end{center}
\vspace{-0.7cm}
\caption{Generalization relationships among program classes.}
\label{fig:gener}\vspace{-5mm}
\end{figure}

\vspace{-7mm}
\paragraph{\bf Weak-Stickiness of the Chase.} Weak-stickiness also has a semantic  version,  called ``weak-chase-stickiness" (WChS); which is implied by  the former.  So as for chase-stickiness, weak-chase-sticky
  programs have a tractable QA problem, even with a possibly infinite chase. This class is one of the two we introduce and investigate. They appear in double-edged boxes in Figure~\ref{fig:gener},
  with dashed edges indicating a semantic class.

By definition, weak-chase-stickiness is obtained by relaxing the condition for ChS: it applies only to values for repeated variables in the body of $\sigma$
that appear in so-called {\em infinite positions}, which are semantically defined. A position is infinite if there is an instance $D$ for which an  unlimited number of different values appear in  $\nit{Chase}(\Sigma,D)$.

Given a program, deciding if a position is infinite is unsolvable, so as deciding in general if the chase terminates. Consequently, it is also undecidable if a program is WChS\ignore{since WChS is defined based on the notion of (in)finite positions}. However, there are syntactic conditions on programs~\cite{FG03,RUD11} that determine some (but not necessarily all) the finite positions. For example, the notion of  position {\em rank}, based on
the program's {\em dependency graph}, are used in~\cite{FG03,AC12} to identify a (sound) set of finite positions, those with {\em finite rank}. Furthermore,
finite-rank  positions are used in~\cite{AC12} to define weakly-sticky (WS) programs as a syntactic subclass of WChS.

\vspace{-2mm}
\paragraph{\bf Finite Positions and Program Classes.} In principle, any set-valued function $S$ that, given a program, returns a subset of the program's finite positions can be used to define a subclass WChS($S$) of WChS.
This is done by applying the definition of WChS above with ``infinite positions" replaced by ``non-$S$-finite positions". Every class WChS($S$) has a tractable QA problem.

$S$ could be computable on the basis of the program syntax or not. In the former case, it would be a ``syntactic class". \ignore{${\it WChS}(.)$ is a mapping that returns a class of programs by replacing the (in)finite positions in the definition of WChS programs with the positions specified by $S$.}
 Class ${\it WChS}(S)$ grows monotonically with $S$ in the sense that if $S_1 \subseteq S_2$ (i.e. $S_1$ always returns a subset of the positions returned by $S_2$), then ${\it WChS}(S_1) \subseteq {\it WChS}(S_2)$.
 In general, the more finite positions are (correctly) identified (and the consequently, the less finite positions are treated as infinite), the more general the subclass of WChS that is identified or characterized.

For example, the function $S^\bot$ that always returns an empty set of finite positions, ${\it WChS}(S^\bot)$ is the class of sticky programs, because stickiness must hold no matter what the (in)finite positions are. At the other extreme, for function $S^\top$ that returns all the (semantically) finite positions, ${\it WChS}(S^\top)$ becomes the class WChS. (As mentioned above, $S^\top$ is in general uncomputable.)  Now, if $S^{\it rank}$  returns the set of finite-rank positions  (for a program $\mc{P}$, usually denoted by $\Pi_F(\mc{P})$ \cite{FG03}), ${\it WChS}(S^{\it rank})$ is the class of WS programs.

\vspace{-2mm}
\paragraph{\bf Joint-Weakly-Stickiness.} The {\em joint-weakly-sticky} (JWS) programs we introduce form a  syntactic class strictly between WS and  WChS. Its definition appeals to the notions of {\em joint-acyclicity} and {\em existential dependency graphs} introduced in~\cite{RUD11}. Figure~\ref{fig:gener} shows this syntactic class, and the inclusion relationships between classes of \dpm programs.\footnote{Rectangles with dotted-edges show semantic classes, and  double-edged rectangles show the classes introduced in this work. Notice that programs in semantic classes include the instance $D$, but syntactic classes are data-independent (for any instance as long as the syntactic conditions apply).}

If $S^{\it ext}$ denotes the function that specifies finite positions on the basis of the {\em existential dependency graphs} (EDG), implicitly defined in~\cite{RUD11}, the JWS class is, by definition, the class ${\it WChS}(S^{\it ext})$. EDGs provide a finer mechanism for capturing (in)finite positions in comparison with positions ranks (defined through dependency graphs): \ $S^{\it rank} \subseteq S^{\it ext}$. Consequently, the class of JWS programs, i.e. ${\it WChS}(S^{\it ext})$, is a strict superclass of WS programs, i.e. ${\it WChS}(S^{\it rank})$.\footnote{The JWS class is different from (and incomparable with) the class of {\em weakly-sticky-join}  programs (WSJ) introduced in~\cite{ACL10}, which extends the one of WS programs with consideration that are different from those used for JWS programs. WSJ generalizes WS on the basis of the weakly-sticky-join property of the chase~\cite{ACL10,AC12} and is related to repeated variables in single atoms.}

\vspace{-2mm}
\paragraph{\bf QAA for WChS.}  Our  query answering algorithm for WChS programs is parameterized by a (sound) finite-position function $S$ as above. It is denoted with ${\it AL}^S$, and  takes as input $\Sigma, D$, query $\mc{Q}$, and $S(\Sigma)$, which is a subset of the program's finite positions (the other are treated as infinite by default).

The customized algorithm ${\it AL}^S$ is guaranteed to be sound and complete only when applied to programs in ${\it WChS}(S)$: \ ${\it AL}^S(\Sigma,D,\mc{Q})$ returns all and only the query answers. (Actually, ${\it AL}^S$ is still sound for any program in WChS.) ${\it AL}^S$ runs in polynomial-time in data; and
 can be applied to both the WS and the JWS syntactic classes. For them the finite-position functions are computable.

${\it AL}^S$ is based on the concepts of {\em parsimonious chase} ({\em pChase}) and {\em freezing nulls}, as used for QA with {\em shy Datalog}, a fragment of \de \cite{LE11}. At a {\em pChase} step, a new atom is added only if a homomorphic atom is not already in the chase. Freezing a null is promoting it to a constant (and keeping it as such in subsequent chase steps). So, it cannot take (other) values under homomorphisms, which may create new {\em pChase} steps. Resumption of the {\em pChase}
means freezing {\em all} nulls, and continuing {\em pChase} until no more {\em pChase} steps are applicable.

Query answering with shy programs has a first phase where the {\em pChase} runs until termination (which it does). In a second phase, the {\em pChase} iteratively resumes for a number of times that depends on the number of distinct $\exists$-variables in the query. This second phase is required to properly deal with joins in the query.
Our  QAA for WChS programs (${\it AL}$) is similar, it has the same two phases, but a {\em pChase} step is modified: after every application of a {\em pChase} step that generates nulls,  the latter that appear in $S$-finite positions are immediately frozen.

\vspace{-2mm}
\paragraph{\bf Magic-Sets Rewriting.} \  It turns out that JWS, as opposed to WS, is closed under the quite general magic-set rewriting method \cite{ceri}
 introduced in \cite{AL12}. As a consequence, ${\it AL}$ can be applied to both the original JWS program and its magic rewriting. (Actually, this also holds for the superclass WChS.)

 \newpage

It can be proved that (our modification of) the magic-sets rewriting method in \cite{AL12} does not change the character of the original finite or infinite positions.
 The specification of (in)finiteness character of positions in magic predicates is not required by ${\it AL}$,  because no new nulls appear in them during the ${\it AL}$ execution. As consequence, the MS method rewriting can be perfectly integrated with our QAA, introducing additional efficiency.

\ignore{
\paragraph{\bf An Alternative QAA.} Specially for WS-\dpm, we also studied a different approach for query answering based on grounding variables. We propose a partial grounding algorithm that given a WS-\dpm program replaces some of the variables in the body of the {\it tgd} rules with constants and null values. That is to say, the partial grounding algorithm removes problematic repeated variables that violate syntactic restriction of sticky programs. Consequently, the result of the partial grounding algorithm on a WS-\dpm program is a simpler sticky \dpm program and therefore fo-rewritability of sticky programs allows us to employ query rewriting for answering conjunctive queries. We prove that this partial grounding algorithm generates polynomially many new {\it tgd} rules with respect to the size of {\it edb}.
}

\vspace{1mm}
\noindent \small {\bf Acknowledgments:} \ We are very grateful to Mario Alviano and the DLV team for providing us with information and support in relation to existential Datalog. We also appreciate
useful conversations with Andrea Cali and Andreas Pieris on Datalog$\pm$, and important  comments from Andrea Cali on an earlier version of this paper.

\end{document}